\documentclass[twocolumn]{aa}  

\usepackage{graphicx}
\usepackage{dblfloatfix}
\usepackage{txfonts}
\usepackage{subfigure}
\usepackage{hyperref}
\begin{document} 
\graphicspath{{figures/}}

   \title{An atomic hydrogen bridge fueling NGC\,4418 with gas from VV\,655}

    %\subtitle{VLA observations reveals VV\,655 connected with NGC\,4418}

      \author{E. Varenius \inst{\ref{inst:chalmers}}
\and
F. Costagliola \inst{\ref{inst:chalmers}}
\and
H.-R. Kl{\"o}ckner \inst{\ref{inst:hrk}}
\and
S. Aalto \inst{\ref{inst:chalmers}}
\and
H. Spoon \inst{\ref{inst:cornell}}
\and
I. Mart{\'i}-Vidal \inst{\ref{inst:chalmers}}
\and
J. E. Conway \inst{\ref{inst:chalmers}}
          }

   \institute{
              Department of Space, Earth and Environment,
              Chalmers University of Technology, 
              Onsala Space Observatory,
              439 92 Onsala, 
              Sweden.
              \email{varenius@chalmers.se}
              \label{inst:chalmers}
             \and
              Max-Planck-Institut f\"ur Radioastronomie, Auf dem H\"ugel 69, 53121 Bonn, Germany
              \label{inst:hrk}
\and 
Cornell Center for Astrophysics and Planetary Science, Space Sciences Building, Ithaca, NY 14853, USA
\label{inst:cornell}
          }

   %\date{Received Month DD, YYYY; accepted Month XX, YYYY}
		  \date{Submitted to Astronomy \& Astrophysics}

% \abstract{}{}{}{}{} 
% 5 {} token are mandatory
 
  \abstract
   { The galaxy NGC\,4418 harbours a compact ($<20$\,pc) core with a very high
bolometric luminosity ($\sim10^{11}$L$_\odot$).  As most of the galaxy's energy
output comes from this small region, it is of interest to determine what
fuels this intense activity. An interaction with the nearby blue irregular
galaxy VV\,655 has been proposed, where gas aquired by NGC\,4418 could trigger
intense star formation and/or black hole accretion in the centre.
   }
   {We aim to constrain the interaction hypothesis by studying neutral hydrogen
structures which could reveal tails and debris connecting NGC\,4418 to the
nearby galaxy VV\,655.}
   {We present observations at 1.4\,GHz with the Very Large Array (VLA) of
radio continuum as well as emission and absorption from atomic hydrogen.
Gaussian distributions are fitted to observed HI emission and absorption
spectra. We estimate the star formation rate (SFR) of NGC\,4418 and VV\,655
from the 1.4\,GHz radio emission and compare with estimates from archival
70\,$\mu$m Herschel observations.
   }
   {An atomic HI bridge is seen in emission, connecting NGC\,4418 to the nearby
galaxy VV\,655. While NGC\,4418 is bright in continuum emission and seen in HI
absorption, VV\,655 is barely detected in the continuum but show bright HI
emission (M$_\mathrm{HI}\sim10^9$\,M$_\odot$).  We estimate SFRs from the
1.4\,GHz continuum of $3.2$\,M$_\odot$\,yr$^{-1}$ and
$0.13$\,M$_\odot$\,yr$^{-1}$ for NGC\,4418 and VV\,655 respectively. Systemic
HI velocities of $2202\pm20$\,km\,s$^{-1}$ (emission) and
$2105.4\pm10$\,km\,s$^{-1}$ (absorption) are measured for VV\,655 and NGC\,4418
respectively.  Redshifted HI absorption is seen
($v_c=2194.0\pm4.4$\,km\,s$^{-1}$) towards NGC\,4418, suggesting gas infall.
North-west of NGC\,4418 we detect HI in emission, blueshifted
($v_c=2061.9\pm5.1$\,km\,s$^{-1}$) with respect to NGC\,4418, consistent with an
outflow perpendicular to the galaxy disk.  We measure a deprojected outflow
speed of 178\,km\,s$^{-1}$ which, assuming simple cylindrical model, gives an
order of magnitude estimate of the HI mass outflow rate of
$2.5$\,M$_\odot$\,yr$^{-1}$.
   }
   {
The morphology and velocity structure seen in HI is consistent with an
interaction scenario, where gas was transferred from VV\,655 to NGC\,4418.
Some gas is falling towards NGC\,4418, and may fuel the activity in the centre.
We interpret blueshifted HI-emission north-west of NGC\,4418 as a continuation
of the outflow previously discussed by \cite{sakamoto2013}, powered by star
formation and/or black hole accretion in the centre. 
   }


   \maketitle
%
%________________________________________________________________

\section{Introduction}
The galaxy NGC\,4418, at a distance of 34\,Mpc, has a nucleus which is very
bright in the infrared, but emits significantly less radio emission than
expected from the FIR-radio correlation for star forming galaxies
\citep{yun2001}.  Since \cite{roche1986}, multiple studies have presented
observational evidence compatible with both a radio-weak AGN and/or a young
starburst (see \cite{sakamoto2013}, \cite{costagliola2013} and references
therein). 

Redshifted absorption lines have been found towards the galaxy's nucleus, for
instance in the far-infrared OI and OH lines by \cite{gonzales2012} with
\emph{Herschel}\footnote{Herschel is an ESA space observatory with science
instruments provided by European-led Principal Investigator consortia and with
important participation from NASA.} , and in 21\,cm by \cite{costagliola2013}
with the Multi-Element Radio Linked Interferometer Network (MERLIN) indicating
infall of molecular and atomic gas. \cite{sakamoto2013} also report a U-shaped
feature visible in the optical extending towards the north-west, i.e. along the
rotation axis of the disk, which they interpret as an outflow.

The NGC\,4418 nucleus contains a Compton thick compact core ($<20$\,pc)
surrounded by molecular gas \citep{sakamoto2013}.  The nucleus is deepely
obscured, evidenced by the deep 9.7\,$\mu$m silicate absorption feature
\citep{roche1986,spoon2001} and the overall resemblance to an embedded
protostar in the mid-IR \citep{spoon2001}.  If a starburst is powering the
source, it is very deficient in PAH emission, likely caused by extreme
conditions and extinction in the obscuring medium around the star forming
regions.  The H$\alpha$ emission from NGC\,4418 is limited to the nucleus
\citep[][their Fig 5, page 132]{lehnert1995}, suggesting that the rest of the
galaxy either lacks OB stars capable of exciting HII regions, or that HII gas
in the disk is absent.

ALMA observations show a very rich spectrum of molecules in an environment
similar to the nuclei of the extreme starburst galaxy Arp\,220
\citep{costagliola2015}.  \cite{costagliola2013} argue that if a central
starburst is responsible for all the energy output, it must be 3-10\,Myrs old
and have a star formation rate $>10$\,M$_\odot$ yr$^{-1}$.  The complex radio
morphology reported by \cite{varenius2014} indeed argues in favour of a
significant starburst component. If this is the case, what triggered this
recent starburst in the centre of NGC\,4418?

\cite{roche1986} noted that NGC\,4418 may be interacting with the nearby galaxy
VV\,655 (also known as MCG+00-32-013) about $3'$ South-East of NGC\,4418.
VV\,655 has blue colors, irregular structure and emission lines consistent with
vigorous star formation \citep{2003astro.ph..6581C, sdss12}.
Although this possible companion galaxy is shown by \cite{kawara1990}, their
Fig. 1, and mentioned by \cite{evans2003} and \cite{costagliola2013}, no clear
evidence has been presented for any interaction between NGC\,4418 and VV\,655.  

In this paper we present radio observations taken with the VLA to search for
neutral hydrogen structures connecting NGC\,4418 to VV\,655.  The paper is
organised as follows. In Sect.  \ref{sect:obs} we describe the observations and
data reduction. In Sect.  \ref{sect:results} we present the results of analysis
and modeling done using the data. The results are discussed in Sect.
\ref{sect:discussion} and finally summarised in Sect.  \ref{sect:summary}.  In
this paper we assume a distance to NGC\,4418 (and VV\,655) of 34\,Mpc, i.e.
$1''=161$\,pc, as given by \cite{sakamoto2013}. 

%For reference, VV\,655 and NGC\,4418 are listed with redshifts of 0.007346148
%and z=0.007084649 respectively in SDSS DR13 \footnote{See e.g.
%http://skyserver.sdss.org/dr13/en/tools/quicklook/summary.aspx} which,
%according to \cite{wright2006}, imply luminosity distances of 31.8\,Mpc and
%30.7\,Mpc. However, 
%In this paper we assume a distance to NGC\,4418 (and VV\,655)
%of 34\,Mpc, i.e.  $1''=161$\,pc, as given by \cite{sakamoto2013}. 

\section{Observations and data reduction}
\label{sect:obs}
We present data from the VLA taken in D configuration in
project AS748 (PI: H. Spoon) on 2003-03-21 with 5.3h on source. The parrallel
hands (RR,LL) were correlated with phase-centre R.A.  $12^h26^m59^s.750$ and
Dec.  $-00^\circ53'31.50''$ and the data stored in the archive with time
resolution of 30 seconds and spectral resolution of 48.8\,kHz per channel.  The
total bandwidth covered 63 channels, centered on 1.41017559\,GHz.  

The data were edited and calibrated in a standard manner using Astronomical
Image Processing System (AIPS) \citep{greisen2003} version 31DEC15. The
Python-based interface to AIPS, ParselTongue \citep{kettenis2006}, was used to
script the calibration. The flux scale and bandpass was calibrated using 3C286,
which was assumed to be 15.1025\,Jy at the centre of the band as calculated by 
the AIPS task \verb!SETJY!.  The visibility phases were
referenced to the calibrator 1150-003 at R.A. $11^h50^m43.8707^s$
$-00^\circ23'54.204''$ \footnote{Within 1 milliarcsecond of the current catalog
position at http://astrogeo.org/calib/search.html.}, found to be 2.94\,Jy.  The
corrections derived from 3C286 and 1150-003 were applied to NGC4418.  Visual
inspection revealed HI-emission in the channel range 16-53. The remaining
channels were used to fit and remove the continuum in the UV-domain using the
task \verb!UVLSF! in AIPS. 

The continuum-subtracted data were imaged using the task \verb!CLEAN!  in CASA
4.6 with Briggs weighting (robust=1) using baselines $>300\lambda$ (to
filter out far-away interferring sources).  During
cleaning, the data were gridded in velocity (using the optical velocity
definition), and CLEAN corrected for effects of the Earth's rotation
and movement with respect to the solarsystem barycentre, assuming a HI
rest-frequency of 1420.405751\,MHz.  A primary-beam corrected image cube was
obtained of the stokes I emission with image pixel size $4''$.  The spectral
cube had a velocity resolution of $10$\,km/s and RMS noise of
$\sim$0.4\,mJy\,beam$^{-1}$ in each velocity channel.  A common synthesized
CLEAN restoring beam of $61.5''\times50.6''$ with position angle (PA)
$-1.19^\circ$ was used for all velocity channels.  

We also obtained a 1.4\,GHz continuum image by deconvolving
the line-free output from \verb!UVLSF! using the Multi-Frequency Synthesis
gridding offered by \verb!CLEAN!. The continuum image RMS noise was
0.2\,mJy\,beam$^{-1}$ with a synthesized beam of $61.5''\times51.0''$ with PA
$-1.34^\circ$.  

To get an independent estimate of the SFR we checked the 70\,$\mu$m flux
densities of NGC\,4418 and VV\,655  as observed by Herschel
PACS \citep{2010A&A...518L...1P,2010A&A...518L...2P} using data from the
Herschel Science Archive.  We used data taken 2011-07-17 towards NGC\,4418 with
scan id 1342224340, available as a FITS image (with calibration level 2.5).

%http://astrogeo.org/cgi-bin/calib_search_form.csh?ra=+11%3A50%3A43.8707&dec=-00%3A23%3A54.204&num_sou=5&format=html

\section{Results}
\label{sect:results}
In this section we present the results of the analysis of the continuum and
spectral line images. Note that both images are available online via the
CDS\footnote{See CDS URL}.

\subsection{1.4\,GHz continuum}
\label{sect:continuum}
\begin{figure}[htbp]
\centering
\subfigure[1.4 GHz continuum]{
        \includegraphics[width=0.45\textwidth]{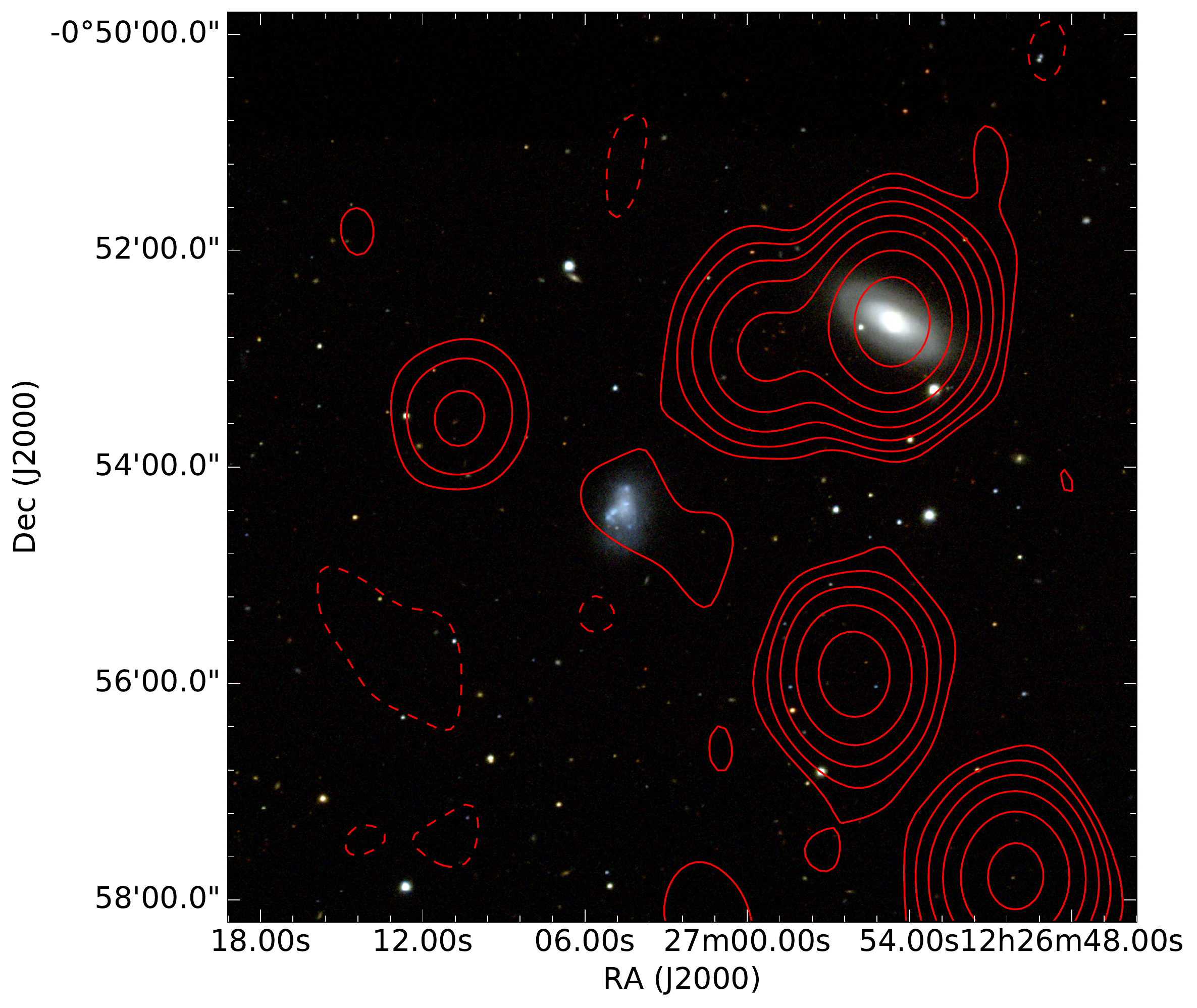}
        \label{fig:continuum}
}
\subfigure[HI moment 0 ]{
        \includegraphics[width=0.45\textwidth]{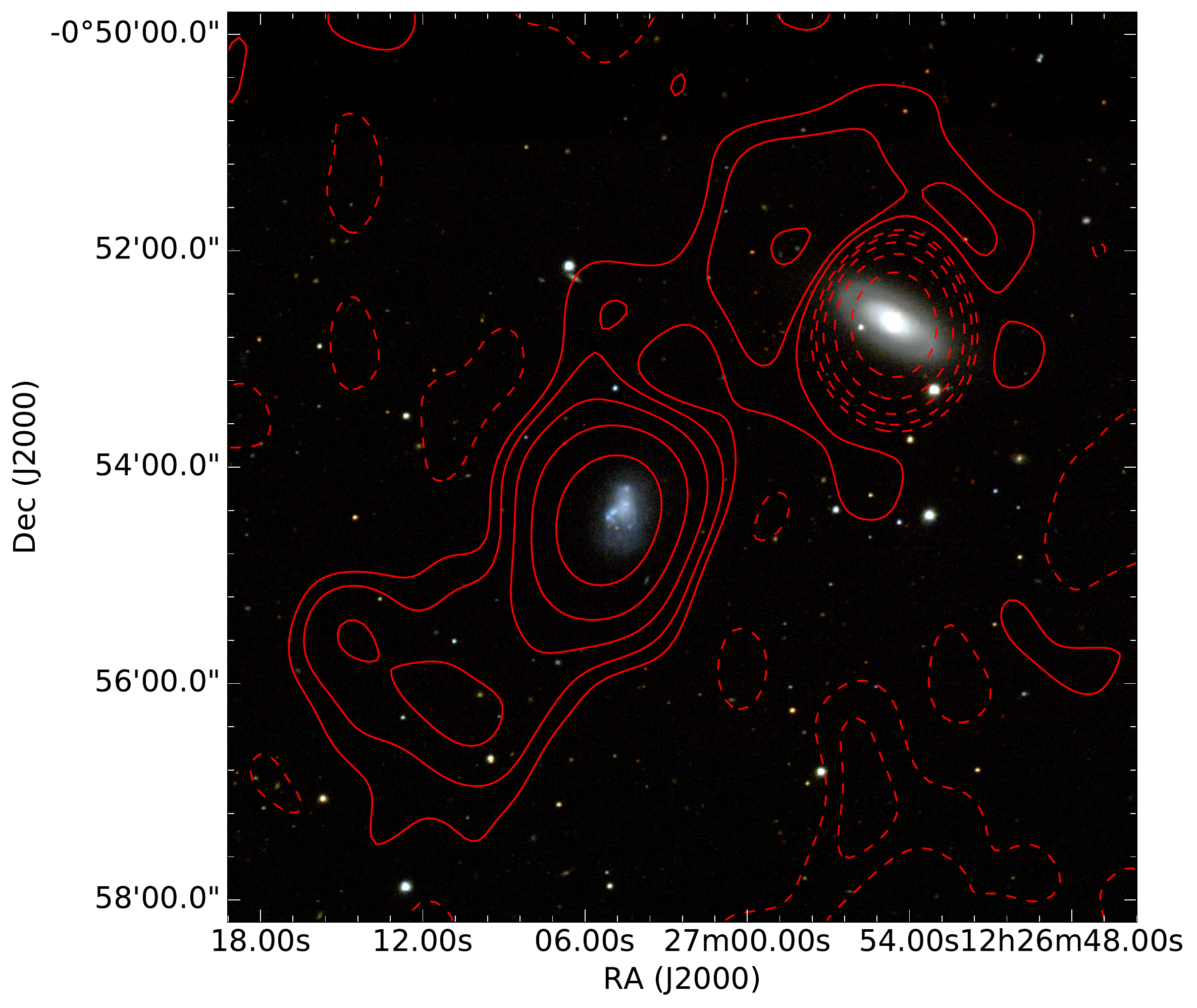}
        \label{fig:moment0}
}
\caption{
Panel \subref{fig:continuum} shows the 1.4\,GHz continuum plotted as contours
of $(-2,2,4,8,16,32,64,128)\times0.2$\,mJy\,beam$^{-1}$ overlayed on a three
colour SDSS image (using i,r,g bands). NGC\,4418 is the white disk in the
top-right part of the image, and VV\,655 is the blue irregular galaxy near the
image centre.  In addition to NGC\,4418, four other sources are visible above
$4\sigma$ as noted in Sect.\ref{sect:continuum}.  VV\,655 is barely detected at
$3\sigma$ in the radio continuum. Panel \subref{fig:moment0} shows the moment 0
map of the continuum-subtracted HI (in emission and absorption) with contours
at ($-32,-16,-8,-4,-2,2,4,8,16,32$)$\times$35\,mJy\,beam$^{-1}$\,km\,s$^{-1}$.
}
\end{figure}
The 1.4\,GHz continuum is presented in Fig. \ref{fig:continuum} as contours
overlayed on a three colour (i,r,g-band) image made from the Sloan Digital Sky
Survey (SDSS) \footnote{Using data from SDSS data release 12 \citep{sdss12}
available via http://dr12.sdss3.org/mosaics/.}. 
In addition to NGC\,4418, four other sources are visible above $4\sigma$ in
Fig. \ref{fig:continuum} which are all identified as compact ($<5''$) sources
with counterparts in the FIRST Survey \citep{FIRST1995} with the same flux
density as measured in this work.  Unfortunately we did not find any redshifts
for these sources which could prove that they are background sources. However,
given that they are compact, having no clear relation to either NGC\,4418 or
VV\,655 (i.e. no extensions or alignments which could suggest e.g. AGN jet hot
spots), and do not show any HI emission in Fig. \ref{fig:moment0} (i.e.
indicate significantly different redshift) we assume these are background
sources.

Assuming the eastern extension is a background source, a fit of a Gaussian
intensity distribution to the image finds NGC\,4418 to be unresolved with an
integrated flux density of 40\,mJy, in good agreement with the 38.5\,mJy measured
by \cite{baan2006} using the VLA.  As noted by \cite{costagliola2013}, this is
also in agreement with the 38\,mJy measured with MERLIN, which indicates that
there is no significant 1.4\,GHz emission on scales larger than the $0.5''$
component sampled by MERLIN (\cite{costagliola2013}, their Fig. A.1).

\subsubsection{Star formation rate estimates}
The galaxy VV\,655 is barely detected at the $3\sigma$ level
in our 1.4\,GHz continuum image (Fig. \ref{fig:continuum}) i.e.
L$_{1.4\mathrm{GHz}}\sim8.2\times10^{19}$\,W\,Hz$^{-1}$. 
%4*pi*(34 Mpc)**2*1e-26*0.6e-3 (W/m**2/Hz) in (W/Hz) = 8.29889059e19 W / Hz
Using Eq. 6 by \cite{bell2003} this corresponds to a star formation rate (SFR) of
%5.52e-22*8.29889059e19/(0.1+0.9*(8.29889059e19/6.4e21)**0.3)
$0.13$\,M$_\odot$\,yr$^{-1}$, although \cite{bell2003} states uncertainties as
large as a factor of 5 for galaxies of this low luminosity. For comparison,
using the same equation given the 1.4\,GHz flux density of 40\,mJy
% 4*pi*(34 Mpc)**2*1e-26*40e-3 (W/m**2/Hz) in (W/Hz)
(5.53$\times10^{21}$\,W\,Hz$^{-1}$) we obtain an integrated radio SFR of 
%5.52e-22*5.53e21/(0.1+0.9*(5.53e21/6.4e21)**0.3)
$3.2$\,M$_\odot$\,yr$^{-1}$ for NGC\,4418.

From the Herschel 70\,um image we measure 125\,mJy associated with VV\,655
which, assuming the same distance to VV\,655 as for NGC\,4418 of 34\,Mpc,
translates to $L(70\mu$m$)=7.39\times10^{41}$\,erg\,s$^{-1}$.
\cite{2010ApJ...714.1256C} present a relation to estimate SFR from 70\,$\mu$m
emission. While they do not claim accurate SFR estimates for individual
galaxies, we use their Eq. 22 to get an order of magnitude estimate for VV\,655
and we obtain SFR(70\,$\mu$m)$\sim0.0435$\,M$_\odot$yr$^{-1}$, i.e. within
a factor of 3 from the 1.4\,GHz estimate.  We note, however, that
VV\,655 is below the validity limit of $1.4\times10^{42}$\,erg\,s$^{-1}$ stated
by \cite{2010ApJ...714.1256C} and therefore this value should be interpreted
with caution.  For comparion, using the same equation with the 39.3\,Jy
($2.33\times10^{44}$\,erg\,s$^{-1}$) measured at 70\,$\mu$m for NGC\,4418 we
estimate SFR(70\,$\mu$m)$\sim13.7$\,M$_\odot$yr$^{-1}$.  NGC\,4418 is known to
be radio weak relative to the FIR-radio correlation \citep{yun2001} so
SFR(70\,$\mu$m)$>$SFR(1.4\,GHz) is expected.  

\subsection{HI in emission and absorption}
\label{sect:HI}
Fig. \ref{fig:moment0} shows the HI emission and absorption as a moment 0 map,
produced by the task \verb!immoments! in CASA. NGC\,4418 is surrounded by HI
seen in emission outside about $1'$ distance from the centre, and seen in
absorption inside this radius (see Fig.  \ref{fig:moment0}). HI emission is
prominent towards VV\,655 and in regions between the two galaxies.  
The moment 0 peak of absorption is $-52$\,mJy\,beam$^{-1}$km\,s$^{-1}$,
towards the centre of NGC\,4418 and the peak of emission is
2.0\,Jy\,beam$^{-1}$\,km\,s$^{-1}$, towards centre of VV\,655.  A
bridge of emission seem to connect VV\,655 and NGC\,4418, although this
projection does not prove a physical connection - see further Sect.
\ref{sect:3dproj}.  Finally, there is prominent emission south-east of VV\,655
with a peak of 0.34\,Jy\,beam$^{-1}$km\,s$^{-1}$.

\begin{figure*}[htbp]
\centering
\subfigure[HI moment 1 map with moment 0 contours]{
\includegraphics[height=0.27\textheight]{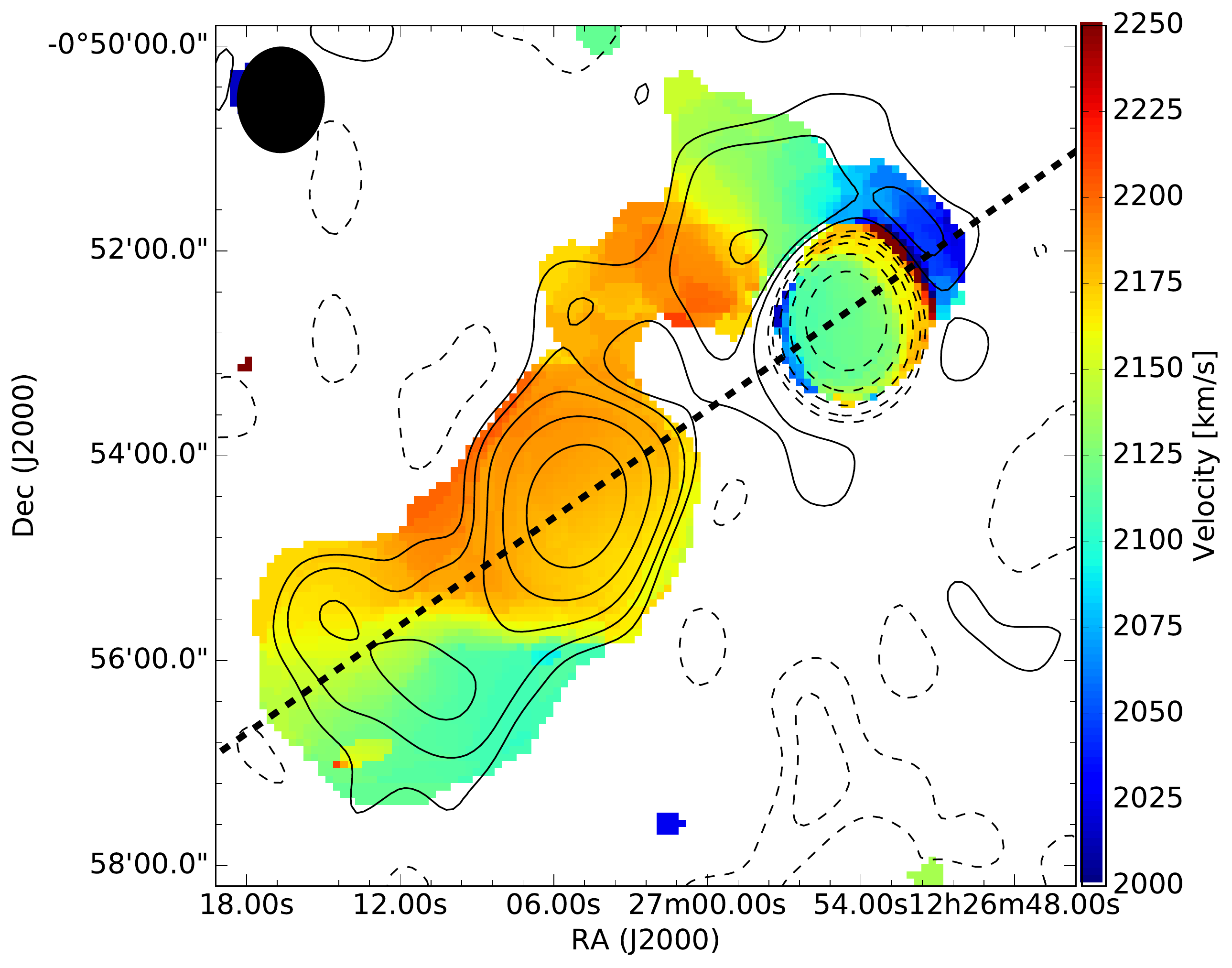}
\label{fig:moment1}
}
\subfigure[PV-diagram along the dashed line in panel \subref{fig:moment1}.]{
\includegraphics[height = 0.27\textheight]{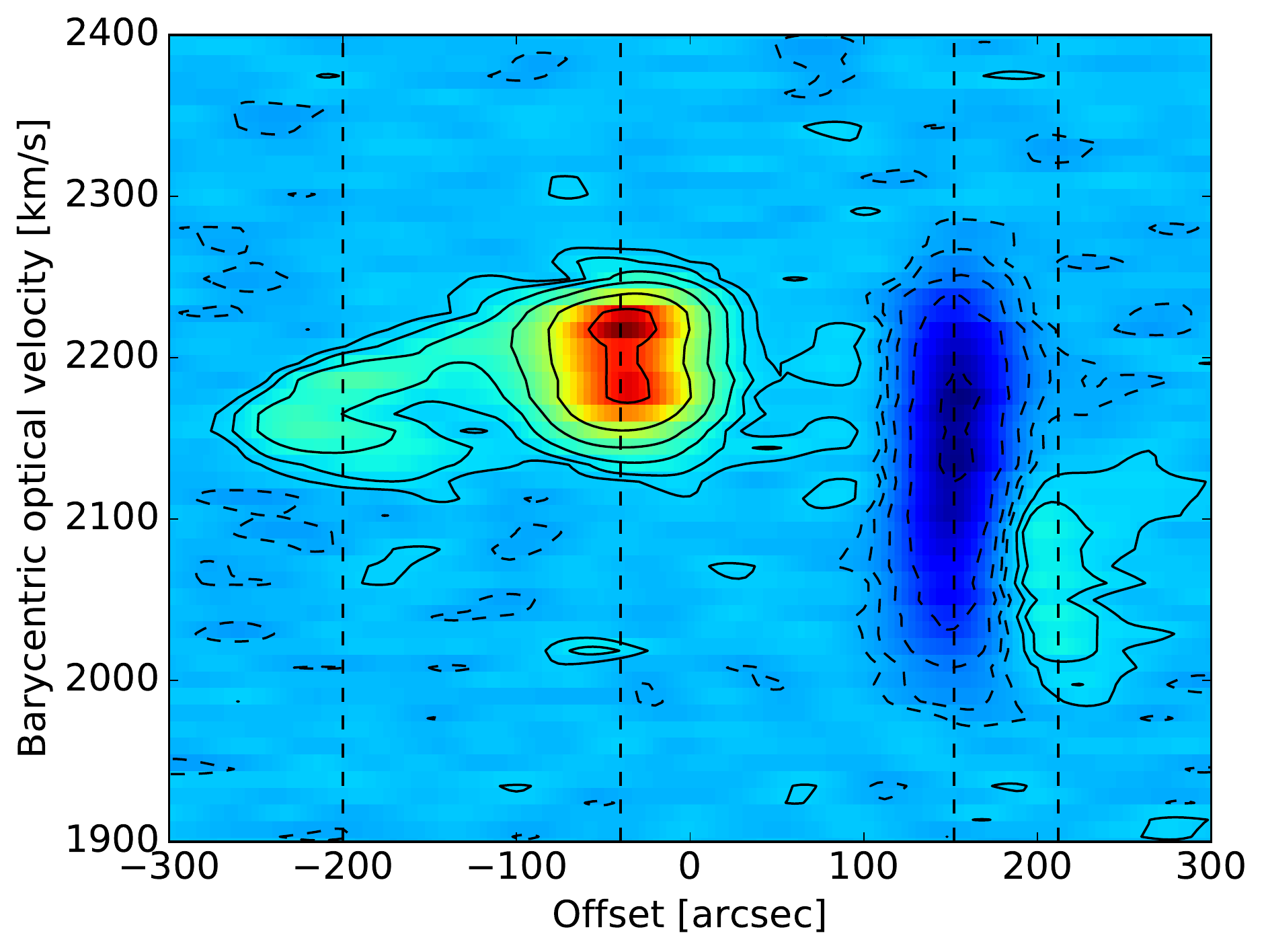}
\label{fig:pv}
}
\caption{
Panel \subref{fig:moment1} shows the moment 1 HI map in color with the moment 0
contours from Fig. \ref{fig:moment0} in black. The CLEAN beam is displayed as a
black ellipse in the top left corner.  Panel \subref{fig:pv} shows a
Position-Velocity diagram extracted along the dashed line in panel
\subref{fig:moment1}; an $11.3'$ slice from R.A.  $12^h27^m21^s.547$ Dec.
$-00^\circ57'19.85''$ to R.A.  $12^h26^m44^s.380$ Dec.  $-00^\circ50'48.73''$,
i.e. going through both NGC\,4418 and VV\,655. Four distinct features can be
seen along the slice at the approximate relative offset values of
$-200''$,$-40''$ (VV\,655), $152''$ (NGC\,4418; in absorption), and $212''$.
Spectra extracted at these positions, i.e. the dashed lines in panel
\subref{fig:pv}, are presented in Fig. \ref{fig:spectra}. Contours in panel
\subref{fig:pv} are $\pm[0.025, 0.05, 0.1, 0.2, 0.4, 0.8]\times$ 26.7\,mJy/beam
(the peak).  \label{fig:mom1andpv}
}
\end{figure*}

The velocity structure of the gas is shown in projection as a moment 1 map in
Fig.  \ref{fig:moment1}, with moment 0 contours in black for easy comparison.
The moment 1 map was obtained after first blanking pixels in the
spectral cube with absolut value smaller than 1.6\,mJy\,beam$^{-1}$. This corresponds
to selecting emission and absorption stronger than $4\sigma$, and was done to
reduce influence from noise which is present also at velocities where there is no
real emission or absorption.  Although the spatial and spectral resolution is
limited, a range of velocities are clearly detected from about 2000\,km\,s$^{-1}$ to
2250\,km\,s$^{-1}$.  This corresponds to emission both red- and blueshifted with respect
to the velocity of 2088\,km\,s$^{-1}$ measured for NGC\,4418 by \cite{sakamoto2013}.

\subsubsection{Position-velocity diagram and spectra}
\label{sect:pv}
\begin{figure}[htbp]
\centering
\includegraphics[width = 0.49\textwidth]{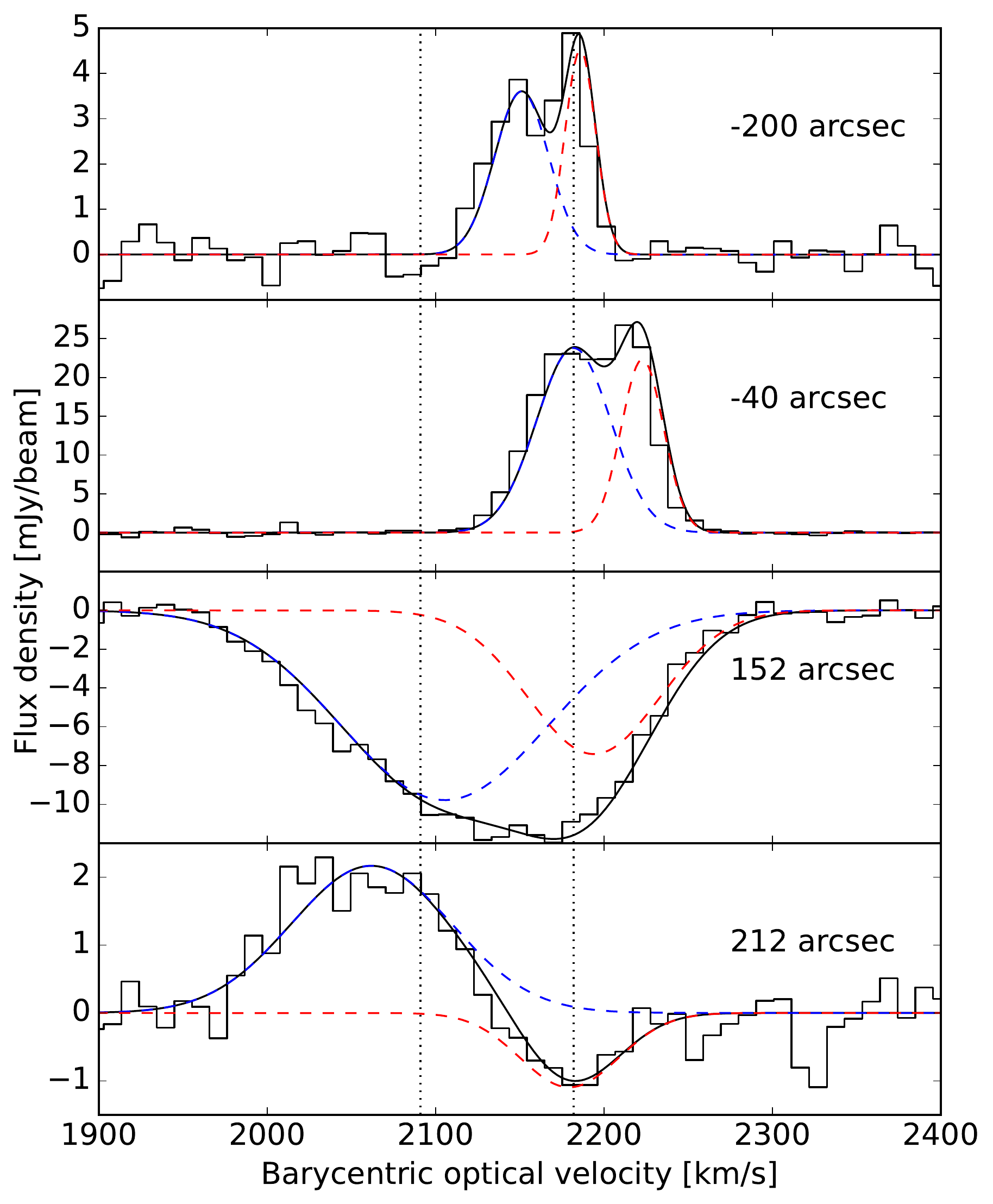}
\caption{Spectra extracted towards the four directions defined in Fig.
\ref{fig:pv}. The dashed curves show the fitted Gaussian components
listed in Table \ref{tab:fitres}. The solid line in each panel represents the
sum of the fitted components, i.e.  to be compared with the observed spectra in
black steps. The vertical dotted lines show the velocities for the two
components fitted to CO(2-1) by \cite{costagliola2013}. For reference, the
systemic velocity of NGC\,4418 has been measured as 2088\,km/s by
\cite{sakamoto2013}.
}
\label{fig:spectra}
\end{figure}

Fig. \ref{fig:pv} shows a position-velocity (PV) diagram extracted from the
spectral cube along the dashed line shown in Fig. \ref{fig:moment1}.  Four distinct
features can be seen at the approximate relative offset positions of
$-200''$,$-40''$ (VV\,655), $152''$ (NGC\,4418; in absorption), and $212''$.
Spectra extracted at these positions, i.e. the dashed lines in panel
\subref{fig:pv}, are presented in Fig. \ref{fig:spectra}. 

The four spectra were modelled using two Gaussian intensity distributions each.
The best fitted parameters are listed in Table \ref{tab:fitres}. Although
limited by spectral and spatial resolution, we note a tentative double
structure in the first two panels (offsets $-200''$ and $-40''$). We interpret
the average HI velocity in the second panel towards VV\,655, of
$2202\pm20$\,km\,s$^{-1}$, as the systemic velocity of the galaxy where the uncertainty
on the central velocity is chosen to cover the two tentative peaks in the spectrum.  The third
panel, towards NGC\,4418, is best fitted with two absorption components. While
the component of \cite{costagliola2013} fitted at $2094\pm22$\,km\,s$^{-1}$ is
in excellent agreement with our result, their second HI component at
$2153\pm11$\,km\,s$^{-1}$ has significantly lower velocity than the one we
find. This may be due to fitting constraints employed by
\cite{costagliola2013}, who did not fit the two components simultaneously.  The
fourth panel corresponds to a region north-west of NGC\,4418.  In addition to
redshifted absorption, we also see prominent blueshifted emission.

Assuming the HI absorption from the 1.4\,GHz continuum of $40$\,mJy dominates
over HI emission towards NGC\,4418, we can estimate the optical depth 
%following Eq. 13.10 by \cite{tools} 
as $\tau\approx-\ln(1+p/40\mathrm{mJy})$ where $p$ are the (negative) peak
values listed in Table \ref{tab:fitres}. The estimated $\tau$ for the two
absorption components towards NGC\,4418 are listed in Table \ref{tab:fitres}.  

\begin{table}[]
\centering
\caption{Best fit parameters for the spectra shown in Fig. \ref{fig:spectra}.}
\label{tab:fitres}
\begin{tabular}{r|r|r|r|r}
Offset & $v_c$ & Peak & FWHM & $\tau_p$\tablefootmark{a}\\
$['']$ & [km/s] & [mJy/beam] & [km/s] & \\
\hline
$-200$ & 2151.1$\pm$2.3 & 3.6$\pm$0.3 & 37.7$\pm$5.4&-\\
$-200$ & 2186.0$\pm$1.2 & 4.5$\pm$0.4 & 21.3$\pm$2.5&-\\
\hline
$-40$\tablefootmark{b} & 2182.0$\pm$0.7 & 23.8$\pm$0.2 & 51.9$\pm$1.3&-\\
$-40$\tablefootmark{b} & 2222.8$\pm$0.4 & 22.3$\pm$0.6 & 29.8$\pm$0.8&-\\
\hline
152\tablefootmark{c} & 2105.4$\pm$10.0 & $-9.8\pm$0.6 & 145.1$\pm$12.8& 0.28\\
152\tablefootmark{c} & 2194.0$\pm$4.4 & $-7.4\pm$1.6 & 92.2$\pm$8.3 & 0.21\\
\hline
212 & 2061.9$\pm$5.1 & 2.2$\pm$0.2 & 112.0$\pm$13.0&-\\
212 & 2179.4$\pm$7.7 & $-1.1\pm$0.2 & 68.7$\pm$17.5&-\\
%Result for slice pos  -200
%Fit result: 2151.1$\pm$2.3 & 3.6$\pm$0.3 & 37.7$\pm$5.4
%Fit result: 2186.0$\pm$1.2 & 4.5$\pm$0.4 & 21.3$\pm$2.5
%Result for slice pos  -40
%Fit result: 2182.0$\pm$0.7 & 23.8$\pm$0.2 & 51.9$\pm$1.3
%Fit result: 2222.8$\pm$0.4 & 22.3$\pm$0.6 & 29.8$\pm$0.8
%Result for slice pos  150
%Fit result: 2105.4$\pm$10.0 & -9.8$\pm$0.6 & 145.1$\pm$12.8
%tau = - ln(1 + p/40mJy) =  0.28
%Fit result: 2194.0$\pm$4.4 & -7.4$\pm$1.6 & 92.2$\pm$8.3
%tau = - ln(1 + p/40mJy) =  0.205
%Result for slice pos  210
%Fit result: 2061.9$\pm$5.1 & 2.2$\pm$0.2 & 112.0$\pm$13.0
%Fit result: 2179.4$\pm$7.7 & -1.1$\pm$0.2 & 68.7$\pm$17.5
\end{tabular}
\tablefoot{
\tablefoottext{a}{Assuming an unresolved background continuum of 40\,mJy from
NGC\,4418 as noted in Sect. \ref{sect:continuum}.}
\tablefoottext{b}{Towards VV\,655.}
\tablefoottext{c}{Towards NGC\,4418.}}
\vspace{-0.5cm}
\end{table}

\subsubsection{An HI emission bridge between NGC\,4418 and VV655}
\label{sect:3dproj}
\begin{figure}[htbp]
\centering
\includegraphics[trim={0cm 0.0cm 0cm 0cm},clip, width =0.45\textwidth]{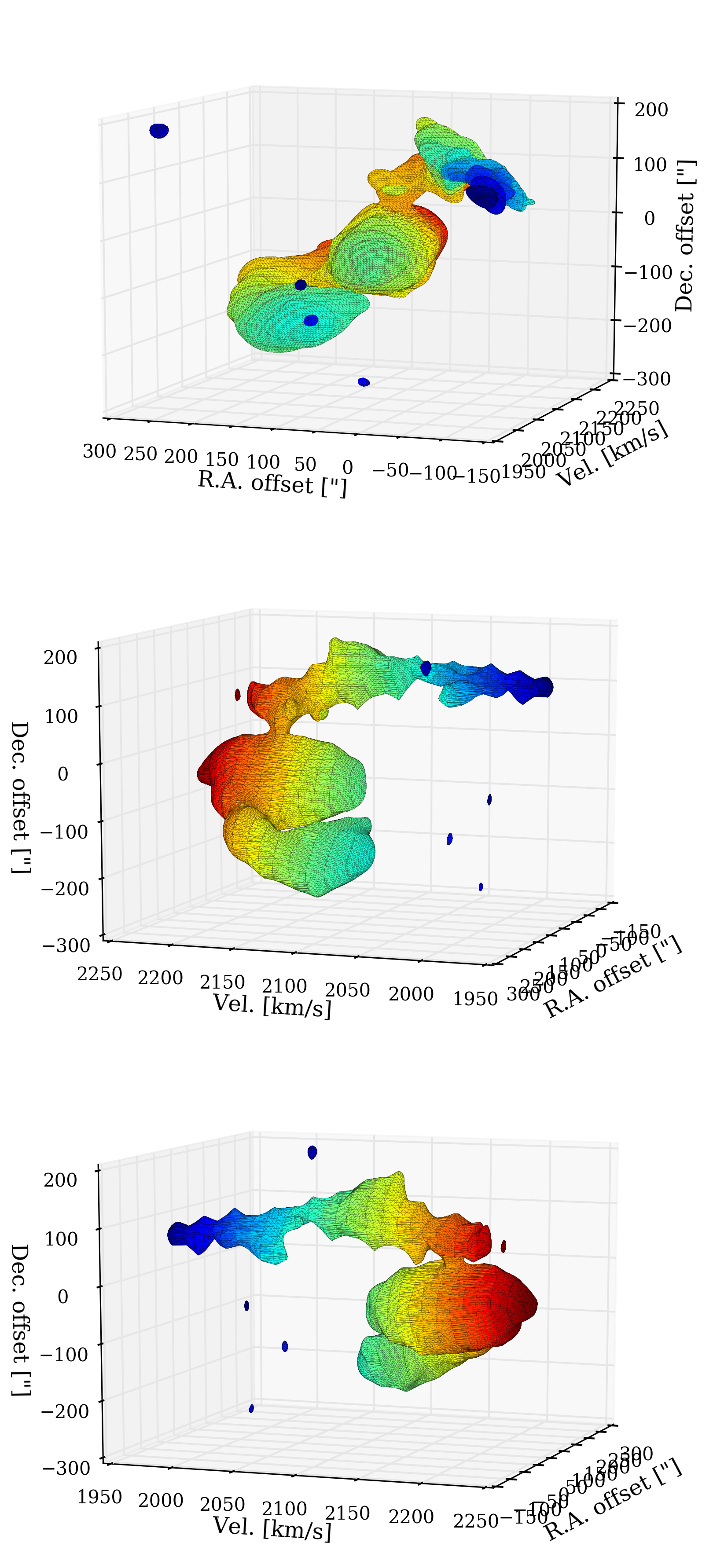} %trim={<left> <lower> <right> <upper>}
\caption{Three projections of the continuum-subtracted HI data cube (R.A.,
Dec., velocity) showing the arc-like connection between NGC\,4418 and VV\,655
in HI emission. For reference, the systemic HI velocities for VV\,655 and
NGC\,4418 are $2202\pm20$\,km\,s$^{-1}$ and $2105.5\pm10$\,km\,s$^{-1}$
respectively (see Sect. \ref{sect:pv}). Note that this figure only shows
the emission, not the absorption towards NGC\,4418.
}
\label{fig:3dproj}
\end{figure}

In Sect. \ref{sect:pv} we noted the systemic HI velocities measured for VV\,655
and NGC\,4418 of $2202\pm20$\,km\,s$^{-1}$ and $2105.5\pm10$\,km\,s$^{-1}$
respectively.  The spectral cube reveals HI-emission connecting VV\,655 and
NGC\,4418. This structure is curved in three-dimensional (R.A., Dec., Velocity)
space and cannot be well illustrated by a moment map (i.e. Figs.
\ref{fig:moment0} and \ref{fig:moment1}) or a single PV diagram (such as Fig.
\ref{fig:pv}).  To show this 3D-structure in two dimensions we plot three
projections of the spectral cube in Fig.  \ref{fig:3dproj}. In addition to the
velocity axis, the velocity information is also reflected in the color map.
Note that the projections only show emission (not absorption), and only pixels
brighter than $4\sigma$ i.e. 1.6\,mJy/beam to reduce influence of noise.  An
animated GIF-image showing more projections are available in the online
material accompanying this paper. From the three-dimensional data it is clear
that there is an HI emission bridge connecting NGC\,4418 and VV\,655.

\subsection{Correction for HI absorption}
\label{sect:4418sub}
As noted in Sect. \ref{sect:continuum}, the 1.4\,GHz continuum from NGC\,4418
appears unresolved in these observations. However, the limited angular
resolution of our data causes absorption towards the nucleus to
blend outwards, reducing nearby emission. To obtain better estimates of the
true HI emission, we therefore corrected for the unresolved absorption in the
cube. This was done by assuming the absorption to be unresolved in each
velocity channel, and therefore completely described by the value of the 
minimum pixel in each channel. To account for noise variations and pixel size
limitations we allowed the minimum position to vary between the channels.  The
variations were limited to the minor beam FWHM of 12 pixels around the
NGC\,4418 position, but the actual shift of the minimum position was less than
3 pixels between all channels.  

\begin{figure}[htbp]
\centering
\includegraphics[width = 0.49\textwidth]{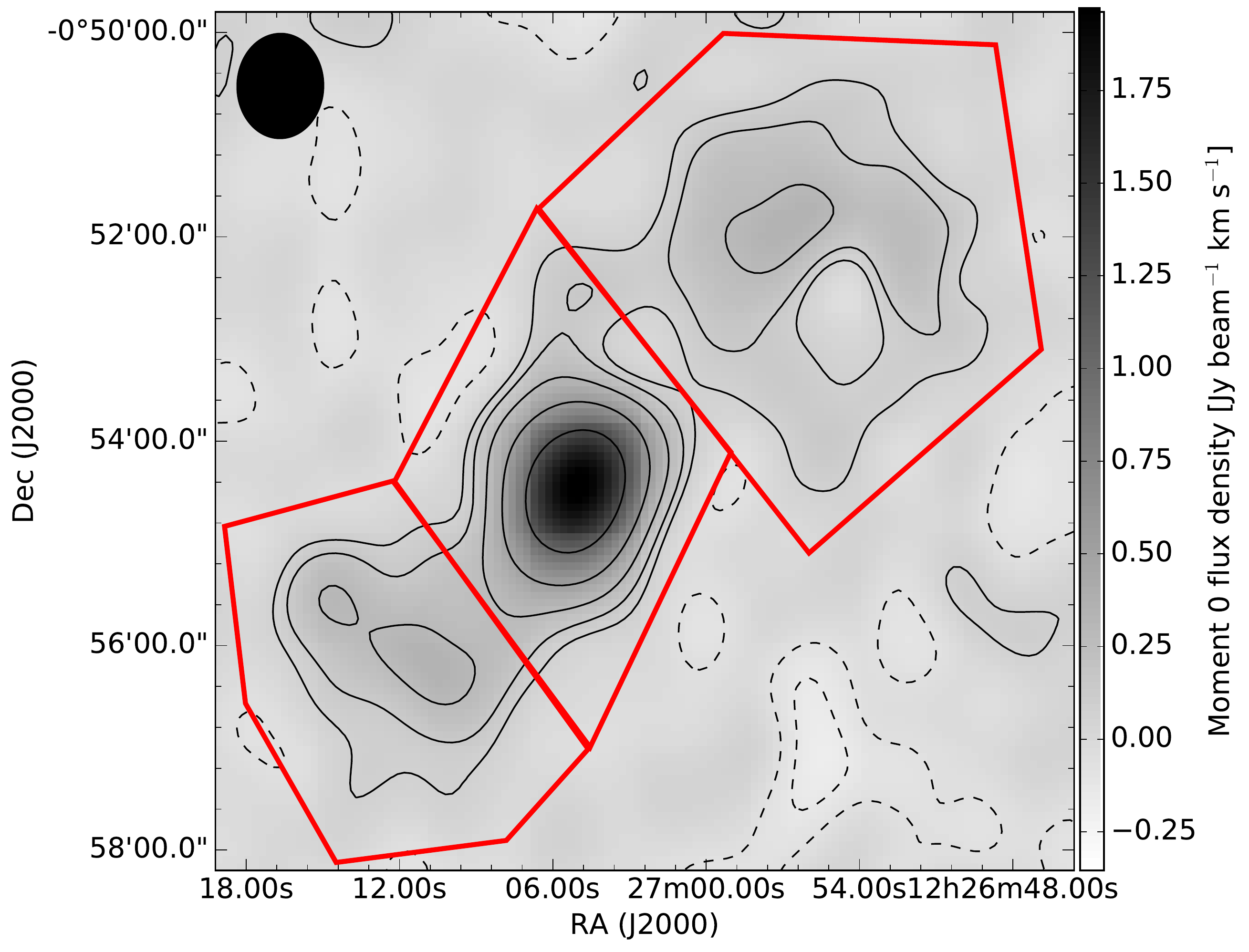}
\caption{Moment 0 map of HI after correcting for the absorption towards
NGC\,4418 as described in Sect. \ref{sect:4418sub}.  More emission is seen
close to NGC\,4418 than in Fig. \ref{fig:moment0}. The red polygons mark three
regions where pixels were summed to measure the flux values listed in Table.
\ref{tab:himass}. Both grayscale and contours show the moment 0 map, with the
same contours as in Fig. \ref{fig:moment0}.
}
\label{fig:4418sub}
\end{figure}

In each velocity channel with negative values towards NGC\,4418, we subtracted
the CLEAN beam multiplied with the minimum pixel value, i.e. assuming an
unresolved background continuum.  The strong absorption feature towards
NGC\,4418 was successfully removed (supporting our single-pixel assumption) and
we now recover more HI emission close to NGC\,4418, see Fig.  \ref{fig:4418sub}
and compare with Fig. \ref{fig:moment0}.  While we restricted our subtraction
to pixels with negative values, there may still be absorption present which
reduce the emission in some channels, although not below zero, i.e the emission
is not optically thin.  Furthermore, we note that our method of correcting for
the absorption cannot recover any HI emission from the nucleus of NGC4418. These
effects may explain the remaining weak depression in the emission towards
NGC\,4418 in Fig.  \ref{fig:4418sub}.

\subsubsection{HI emission fluxes and mass estimates}
To measure the integrated flux density of the HI emission we have divided the
emission in three regions in Fig.  \ref{fig:4418sub}: \emph{North}, defined as
the top-right polygon, includes emission associated with NGC\,4418. The
\emph{Middle} polygon includes emission associated with VV\,655, and the
\emph{South} polygon includes emission associated with the south-east extension
from VV\,655.  The correction for absorption towards NGC\,4418 makes it
possible to estimate the HI gas mass from emission associted also with
NGC\,4418. This value is a lower limit since the emission may not be optically
thin as noted in Sect.  \ref{sect:4418sub}.  The HI emission flux for the three
regions was measured by summing pixels in a moment 0 map, and are given in
Table \ref{tab:himass}.
\begin{table}[]
    \centering
	\caption{Integrated HI emission flux measurements for the three regions
defined in Fig. \ref{fig:4418sub} together with corresponding HI mass estimates
from Eq. \ref{eqn:MHI}. Note that these values were obtained after removing
the absorption towards NGC\,4418 as described in Sect. \ref{sect:4418sub}.
}
\label{tab:himass}
\begin{tabular}{l|r|r}
Region & Flux [Jy\,km\,s$^{-1}$]& M$_\mathrm{HI}$ [M$_\odot$]\\
 \hline
North (NGC\,4418) & 1.47& $4.0\cdot10^8$\\
Middle (VV\,655) & 4.05 &$11.0\cdot10^8$ \\
South & 0.83 &$2.3\cdot10^8$ \\
\hline
Sum & 6.35 & $17.3\cdot10^8$\\
\end{tabular}
\end{table}

From the detected HI emission we estimate the HI mass using Eq. 13.57 in
\cite{tools}:
\begin{equation}
M_{\mathrm{HI}} [\mathrm{M}_\odot]= 2.36\times10^5D^2\int S_\nu d \nu
	\label{eqn:MHI}
\end{equation}
where $\int S_\nu d \nu$ is integrated line flux in Jy\,km/s, and $D$
is the distance in Mpc.  With $D=34$\,Mpc we obtain the values listed in Table
\ref{tab:himass}. We note that the estimated HI mass of VV\,655 is very similar
to the HI masses of other blue compact dwarf galaxies measured by
\cite{thuan2016}.

\section{Discussion}
\label{sect:discussion}
In this section we discuss the results presented in the previous section.

\subsection{NGC\,4418 and VV\,655 are interacting}
The spectral cube presented in Sect. \ref{sect:3dproj} shows that NGC\,4418 and
VV\,655 are connected by a bridge of atomic hydrogen seen in emission.
Furthermore, the similar velocities of the gas surrounding VV\,655 and
NGC\,4418 and the elongated gas morphology can all be explain as signatures of
a recent or ongoing interaction process, where the elongated HI structure
connecting the two galaxies may have formed when the more massive NGC\,4418
disturbed the HI surrounding VV\,655.  We propose that at some time, relatively
recently, NGC\,4418 was gas poor while VV\,655 had a significant HI envelope.
As the galaxies passed each other, the more massive NGC\,4418 stripped away HI
gas from VV\,655. This interaction formed the elongated HI structure we see
today. Indeed, the HI morphology is similar to the structure seen in other
interacting pairs in the \emph{HI Rogues Gallery} \citep{HIroughes}. For example, 
the interacting pair Galaxy Pair II ZW 70/71 (UGC\,9562 and UGC\,9560) 
have an HI emission bridge connecting the two galaxies and an HI extension from
one of the galaxies \citep{2001AJ....121..692C}.

As the HI gas is \emph{only} elongated, and thus the structure is
relatively simple, it is possible that these two galaxies are only passing
once. This could explain the young age of the star burst in the centre of
NGC\,4418, as the swept up gas need time to cool enough to reach the centre of
the galaxy. The tidal forces caused by the interaction can also explain the HI
emission extending south-east of VV\,655.  From the data it is however not
clear if the galaxies are now bound in a merging system or if they are just
passing each other once. Detailed modeling of the velocity structure would
require data with higher spatial and spectral resolution and is beyond the
scope of this paper.

\subsection{HI falling towards NGC\,4418} 
The redshifted absorption detected towards NGC\,4418 (see panel 3 in Fig.
\ref{fig:spectra}) implies gas in front of NGC\,4418 travelling towards the
galaxy.  Indeed, the velocity of about 90\,km/s relative to NGC\,4418 is
similar to the redshifted CO(2-1) component found by \cite{costagliola2013}.
This is consistent with infall, but because of the limited spatial resolution
we cannot probe the detailed dynamics around NGC4418.  While a significant
amount of HI gas is falling towards NGC4418, it is not clear from these data if
this gas does fall all the way into the centre of the galaxy.  However, if some
of the gas does reach the centre, it can explain the recent star formation
and/or AGN activity in the galaxy nucleus. 

\subsection{The north-west outflow} 
\label{sect:outflow}
North-west of NGC\,4418 we detect HI in emission blueshifted with respect to
the systemic velocity of the galaxy. The blueshift can be explained if the
gas is being pushed with an inclination towards us, i.e. out from the galaxy
nucleus.  Indeed, the orientation of this emission is consistent with being a
large-scale continuation of the outflow signature seen by \cite{sakamoto2013},
their Fig. 16.  It is possible that the activity in the nucleus is powering and
outflow which pushes out HI gas surrounding the galaxy. This argument is
supported by the fact that there is no other source visible in this region
which could be the source of this emission.  Such an outflow can be driven by
either intense star formation or AGN activitiy, or a combination of both.
\begin{figure}[htbp]
\centering
\includegraphics[width = 0.49\textwidth]{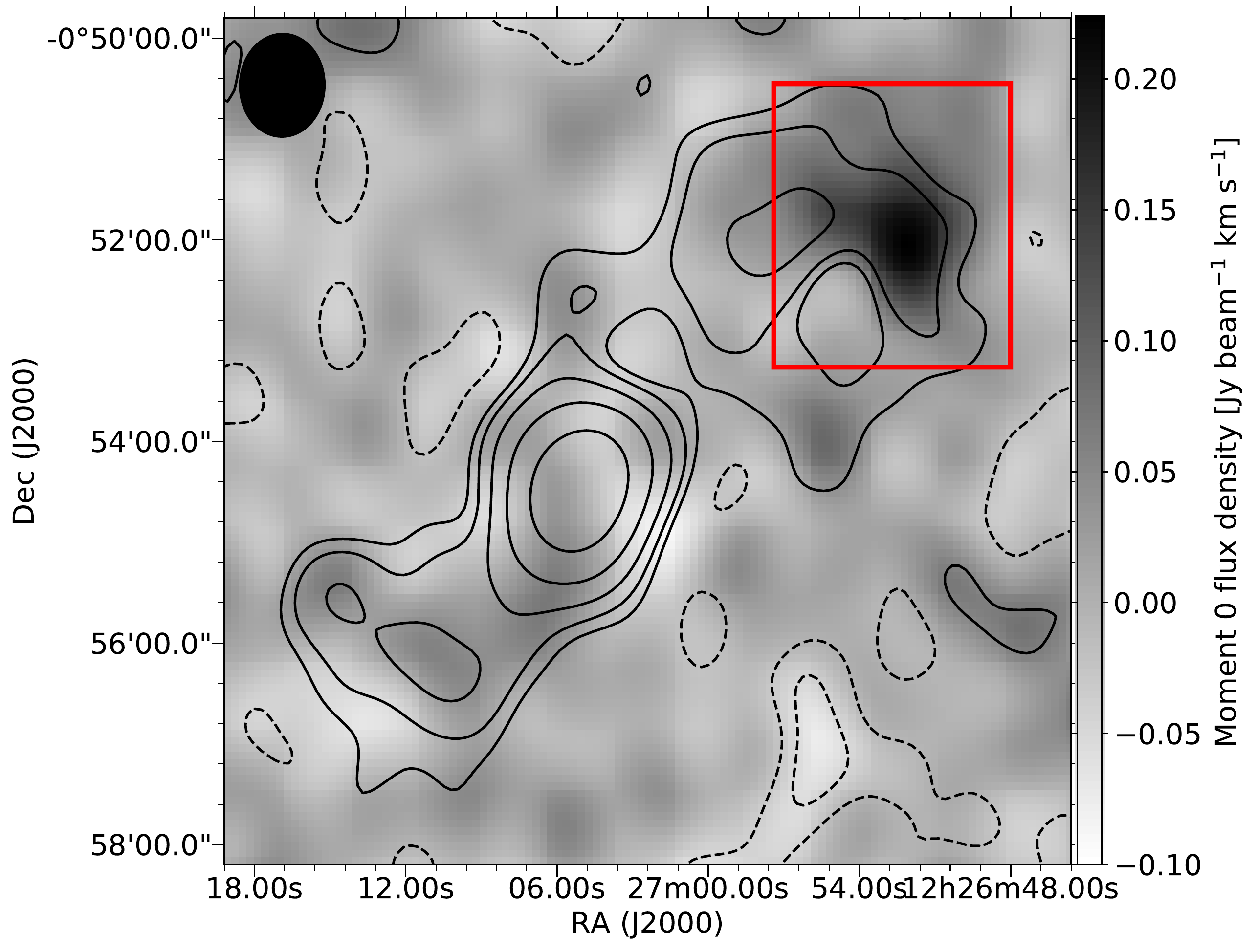}
\caption{In grayscale: moment 0 map of 19 most blueshifted channels, after
first removing the absorption towards NGC\,4418 as described in Sect.
\ref{sect:4418sub}.  Contours are the same as in Fig. \ref{fig:4418sub} to
guide the eye.  The red polygon mark regions where pixels were summed to
measure the flux of the outflow, discussed in Sect.  \ref{sect:outflow}
}
\label{fig:bluemom}
\end{figure}

To measure the HI emission associated with the outflow, we used the
absorption-subtracted image cube (see Sect. \ref{sect:4418sub}) to produce
a moment 0 map, including only the channels containing blueshifted emission
associated with the outflow position. This moment 0 map is shown in grayscale
in Fig. \ref{fig:bluemom}. Summing the pixels in the marked region in Fig. \ref{fig:bluemom}
we measure a flux of 0.48\,Jy\,beam$^{-1}$\,km\,s$^{-1}$ associated with the outflow,
which using Eq. \ref{eqn:MHI} translates to an atomic hydrogen gas mass of
$1.3\times10^{8}$\,M$_\odot$.

We note that the outflow is not clearly resolved in most velocity channels. As
an approximation of the outflow volume, we assume a cylinder with diameter
$25''$ (equal to half the minor axis of the CLEAN restoring PSF) and length
$60''$ (equal to the distance from the centre of NGC\,4418 to the peak of the
outflow emission in Fig. \ref{fig:bluemom}. Given the HI mass estimate above,
this implies an average outflow density of $1.1\times10^{6}$\,M$_\odot$ kpc$^{-3}$.
This is an upper limit, given that the outflow structure is likely conical
instead of a cylinder, as evident from \citealt[][their Fig. 16]{sakamoto2013}.

To estimate the mass outflow rate we need the de-projected outflow velocity. As
an upper limit we take the velocity of the most blueshifted velocity channel
with emission brighter than $3\sigma$, which gives a velocity of
2004\,km\,s${-1}$. (This velocity can also be approximately determined from
Fig. \ref{fig:pv} or the lowest panel in Fig. \ref{fig:spectra}, but using the
full cube provides maximum precision.) Taking the systemic velocity for
NGC\,4418 of 2088\,km\,s$^{-1}$ from \cite{sakamoto2013} implies an outflow
speed of 84\,km/s. Assuming the outflow is perpendicular to the galaxy
disk, which has an inclination angle of 62$^\circ$ \citep{sakamoto2013}, we
obtain a de-projected outflow velocity of 178\,km\,s$^{-1}$. We note that this
is similar to the projected minor-axis velocity shear of 200\,km\,s$^{-1}$
measured in [N$_\text{II}$] as noted by \cite{sakamoto2013}.
% 1.82984431e-7 kpc in one year

Given the above assumptions, we estimate a mass outflow rate of
$2.5$\,M$_\odot$\,yr$^{-1}$. This estimate is to be taken as an order of
magnitude estimate. More accurate constraints require observations with higher
spatial resolution and sensitivity, to properly study the HI velocity structure
on scales between those sampled by MERLIN as presented by
\cite{costagliola2013} and those sampled by VLA as presented in this work.
% 2*2*pi*1.82984431e-7* 1.3e8/(2*2*pi*9.66) = 2.5Msun /year

\section{Summary}
\label{sect:summary}
We present observations at 1.4\,GHz with the VLA of radio
continuum, as well as emission and absorption from atomic hydrogen, towards the
galaxies NGC\,4418 and VV\,655.  We detect a large ($5'\sim50$\,kpc) HI
structure in emission, connecting the two galaxies.  While NGC\,4418 is bright
in continuum emission and seen in HI absorption, VV\,655 is barely detected in
the continuum but show bright HI emission (M$_\mathrm{HI}\sim10^9$\,M$_\odot$).
We estimate SFRs from the 1.4\,GHz continuum of $3.2$\,M$_\odot$\,yr$^{-1}$ and
$0.13$\,M$_\odot$\,yr$^{-1}$ for NGC\,4418 and VV\,655 respectively.  

We fit Gaussian intensity distributions to spectra extracted from the image
cube and measure systemic HI velocities of $2202\pm20$\,km\,s$^{-1}$ (emission)
and $2105.4\pm10$\,km\,s$^{-1}$ (absorption) for VV\,655 and NGC\,4418
respectively.  We also detect redshifted HI absorption
($v_c=2194.0\pm4.4$\,km\,s$^{-1}$) towards NGC\,4418 which suggests gas infall.
The morphology and velocity structure seen in HI is consistent with an
interaction scenario, where gas was transferred from VV\,655 to NGC\,4418.
Some gas is falling towards NGC\,4418, and may fuel the activity in the centre.

North-west of NGC\,4418 we detect HI in emission, blueshifted with respect to
NGC\,4418, consistent with an outflow perpendicular to the galaxy disk.  We
interpret blueshifted HI-emission north-west of NGC\,4418 as a continuation of
the outflow previously discussed by \cite{sakamoto2013}, powered by star
formation and/or black hole accretion in the centre. The observed maximum
velocity of $v_c=2004$\,km\,s$^{-1}$ imples a deprojected speed of
187\,km\,s$^{-1}$. Assuming a steady-state cylindrical outflow from the core of
NGC\,4418, we obtain an order of magnitude estimate of the HI mass outflow rate
as $2.5$\,M$_\odot$\,yr$^{-1}$.

\begin{acknowledgements}
F.C. acknowledges support from Swedish National Research Council grant
637-2013-7261.  This research made use of APLpy, an open-source plotting
package for Python \citep{aplpy2012}.  Using data from the Karl G. Jansky Very
Large Array (VLA), we acknowledge that the National Radio Astronomy Observatory
is a facility of the National Science Foundation operated under cooperative
agreement by Associated Universities, Inc.  Funding for SDSS-III has been
provided by the Alfred P. Sloan Foundation, the Participating Institutions, the
National Science Foundation, and the U.S.  Department of Energy Office of
Science. The SDSS-III web site is http://www.sdss3.org/. The authors
acknowledge assistance from Joachim Wiegert and Niklas Falstad with obtaining
Herschel 70\,$\mu$m SFR estimates. The authors also want to thank John Hibbard
for discussions during the preparation of this manuscript. Finally, the authors
acknowledge the constructive feedback offered by the anonymous referee which
helped to improve the quality of this paper.
\end{acknowledgements}

\bibliographystyle{aa}
\bibliography{../allrefs}

\end{document}